\documentclass[preprint,aps,12pt,showpacs,nofootinbib,tightenlines,superscriptaddress]{revtex4}
\usepackage{verbatim}
\usepackage{amsmath}
\usepackage{graphicx}
\usepackage{amssymb}
\begin{document}
\def\intdk{\int\frac{d^4k}{(2\pi)^4}}
\def\sla{\hspace{-0.17cm}\slash}
\hfill

\title{ Searching for $Z^{'}$ Gauge Boson \\ in an Anomaly-Free U(1)$'$ Gauge Family Model}

\author{Jin-Yan Liu}
\email{jinyanliu@itp.ac.cn}

\affiliation{Kavli Institute for Theoretical Physics China (KITPC) \\
 State Key Laboratory of Theoretical Physics (SKLTP)\\ Institute
of Theoretical Physics, Chinese Academy of Science, Beijing 100190, China }

\author{Yong Tang}
\email{ytang@itp.ac.cn}

\affiliation{Kavli Institute for Theoretical Physics China (KITPC) \\
 State Key Laboratory of Theoretical Physics (SKLTP)\\ Institute
of Theoretical Physics, Chinese Academy of Science, Beijing 100190, China }
\affiliation{Physics Division, National Center for Theoretical Sciences, Hsinchu}

\author{Yue-Liang Wu}
\email{ylwu@itp.ac.cn}

\affiliation{Kavli Institute for Theoretical Physics China (KITPC) \\
 State Key Laboratory of Theoretical Physics (SKLTP)\\ Institute
of Theoretical Physics, Chinese Academy of Science, Beijing 100190, China }

\begin{abstract}
We study a simple ultraviolet(UV) complete and anomaly free $Z'$ model based on a U$(1)'$ gauge family symmetry without introducing extra fermions beyond the standard model. The U$(1)'$ group is diagonal in the three family space in which the U(1)$'$ charges of the first and second families are the same but different from those of the third family. After spontaneous symmetry breaking and rotating to the mass eigenstates of quarks and leptons, there exist in general both flavor-conserving and flavor-changing couplings.  In flavor conserving case,  we use the latest dijet and dilepton search at the LHC to constrain the mass and gauge couplings of $Z'$. Also we show that in the flavor-changing and left-right asymmetry case, the t-channel $Z'$ process can only contribute to part of the forward-backward asymmetry in the top quark pair production observed at the Tevatron by taking into account the constraint from same sign top production at the LHC.

\end{abstract}

\pacs{12.60.cn, 12.60.Fr}

\maketitle
\section{Introduction}

Theoretically, one of the simplest and best motivated extensions of
the standard model(SM) is the introduction of an additional U$(1)'$ gauge symmetry with
the associated gauge boson $Z'$ \cite{Langacker:2008yv}. Such an U$(1)'$ gauge symmetry often appears in
grand unified theories, such as $\textrm{SO}(10)$ or
$\textrm{E}_{6}$ theories. Phenomenologically, as
the extra U$(1)'$ gauge symmetry can result in new interactions between
the new gauge boson and the SM particles, it can lead to some interesting
phenomena or explanations for possible experimental results away from the SM predictions
\cite{Leike:1998wr}.

In gauge theories, the cancelation of anomaly is tightly connected
with renormalizability. In the grand unified theories
\cite{Langacker:1980js} with a large group, such cancelation for U$(1)'$ subgroup can be
 satisfied and new particles if any fill in, automatically. For phenomenological $Z'$ models, with an additional
gauge group introduced, usually the anomaly is canceled with introducing by hand new chiral fermions. Since no new fundamental fermion is observed yet, these new particles have to be heavy and usually difficult to be probed.

From an economical perspective, it is interesting to ask what an $U(1)'$ model looks like if introducing no new chiral fermion. For family universal $U(1)'$ models, unlike family non-universal ones discussed in \cite{Langacker:2000ju, Buchalla:2000sk, Barger:2003hg, Carena:2004xs, Barger:2004hn, King:2005jy, Cheung:2006tm, Barger:2009qs, Chiang:2011cv}, there is no problem like flavor-changing neutral currents(FCNC) effects. However, for such models, there is little room left if no new particles are added as we shall show later. Then we shall not restrict the discussion within family non-universal models and but taking FCNC suppression as an requirement or constraint.

In the present paper, based on the structure of non-linear equations from anomaly cancelation that constrain the $U(1)'$ charges of SM fermions, we shall show a simple UV complete and anomaly free $Z'$ model without introducing any exotic fermions beyond the SM. In this model, the U$(1)'$ charges of the first and second families are the same but different from the third family. A specific feature of this model is that the cancelation of anomalies is satisfied separately in lepton and quark sectors. In general, there can exist both flavor conserving and flavor changing interactions of $Z'$ gauge boson with quarks and leptons after spontaneous symmetry breaking and rotating to the mass eigenstates of quarks and leptons.

Our paper is organized as follows. In Sec.II, we give a brief summary for the anomaly-cancelation conditions of
an extra U(1)$'$ gauge symmetry with the SM particles and the arguments that lead to our model. In Sec. III, we provide a detailed description on our $Z'$ model for the fermion and higgs sectors. In Sec. IV, for flavor-conserving case we use the latest experiments direct search to constrain the parameters of the model and discuss  for flavor-changing case an observable effect, forward-backward asymmetry of top pair production. Our conclusions are given in the last section.

\section{U(1)$'$ anomaly cancelation}

The requirement of vanishing of anomalies is usually used as a guide in building realistic renormalizable theories and many models are then accompanied with new chiral fermions to satisfy the requirement. These models may or may not have a unification origin, but in common they include many particles which is usually too heavy to be detected. For realistic reason, we shall restrict ourselves here to study an anomaly free U(1)$'$ model without introducing any new fermions, and we find a family non-universal model based on simple and reasonable assumptions.

When introducing an additional U(1)$^{\prime}$ gauge symmetry to the SM, we can assign the U(1)$'$ charges $z_{\#}$ to all the SM particles as shown in Table I. In the following discussions, like mentioned above, we shall not introduce any exotic fermions.
\begin{table}[tbph]
 \begin{tabular}{|c|c|c|c|c|}
\hline
 & SU(3)$_{c}$  & SU(2)$_{L}$  & U(1)$_{Y}$  & U(1)$^{\prime}$ \tabularnewline
\hline $Q_{L}^{i}$  & $3$  & $2$  & $+1/3$  & $z_{Q^{i}}$
\tabularnewline \hline $u_{R}^{i*}$  & $\bar{3}$  & $1$  & $-4/3$  &
$z_{u_{i}^{*}}$ \tabularnewline \hline $d_{R}^{i*}$  & $\bar{3}$  &
$1$  & $+2/3$  & $z_{d_{i}^{*}}$ \tabularnewline \hline $L^{i}$  & 1
& $2$  & $-1$  & $z_{L^{i}}$ \tabularnewline \hline $e_{R,i}^{*}$  &
$1$  & $1$  & $+2$  & $z_{e_{i}^{*}}$ \tabularnewline \hline
$\nu_{R,i}^{*}$  & $1$  & $1$  & $0$  & $z_{\nu_{i}^{*}}$
\tabularnewline \hline
\end{tabular}\caption{}
\end{table}
The weak doublets and singlets are denoted as follows,
$\psi=u,d,e,\nu$,
\[
Q_{L}^{i}=\left(\begin{array}{c}
u_{L}^{i}\\
d_{L}^{i}\end{array}\right),\; L^{i}=\left(\begin{array}{c}
\nu_{L}^{i}\\
e_{L}^{i}\end{array}\right),\;\psi_{L,R}^{i}=P_{L,R}\psi^{i},
\]
with $ P_{L}=(1-\gamma_{5})/2,P_{R}=(1+\gamma_{5})/2$.

The anomaly is proportional to the completely symmetric constant
factor,
\[
D_{\alpha\beta\gamma}\equiv\textrm{tr}\left[\left\{
T_{\alpha},T_{\beta}\right\} T_{\gamma}\right]
\]
$T_{\alpha}$ is the representation of the gauge algebra on the set
of all left-handed fermion and anti-fermion fields, and $``\textrm{tr}"$ stands
for summing over those fermion and anti-fermion species. The anomaly free conditions for the theory are given by
\begin{eqnarray}
0 & = &
\sum_{i=1}^{3}(2z_{Q_{i}}+z_{u_{i}^{*}}+z_{d_{i}^{*}}),\quad \left[SU(3)^{2}U(1)^{\prime}\right],\label{eq:3-3-1p}
\nonumber\\
0 & = & \sum_{i=1}^{3}(3z_{Q_{i}}+z_{L_{i}}),\quad \left[SU(2)^{2}U(1)^{\prime}\right],\label{eq:2-2-1p}\nonumber\\
0 & = & \sum_{i=1}^{3}(\frac{1}{6}z_{Q_{i}}+\frac{4}{3}z_{u_{i}^{*}}+\frac{1}{3}z_{d_{i}^{*}}+\frac{1}{2}z_{L_{i}}+z_{e_{i}^{*}}),\quad \left[U(1)_{Y}^{2}U(1)^{\prime}\right],\label{eq:1-1-1p}\nonumber\\
0 & = & \sum_{i=1}^{3}(6z_{Q_{i}}+3z_{u_{i}^{*}}+3z_{d_{i}^{*}}+2z_{L_{i}}+z_{e_{i}^{*}}+z_{\nu_{i}^{*}}),\quad \left[\textrm{global } U(1)^{\prime}\right]\label{eq:g-g-1p}\nonumber\\
0 & = & \sum_{i=1}^{3}(z_{Q_{i}}^{2}-2z_{u_{i}^{*}}^{2}+z_{d_{i}^{*}}^{2}-z_{L_{i}}^{2}+z_{e_{i}^{*}}^{2}),\quad \left[U(1)^{\prime}{}^{2}U(1)_{Y}\right],\label{eq:1p-1p-1}\nonumber\\
0 & = &
\sum_{i=1}^{3}(6z_{Q_{i}}^{3}+3z_{u_{i}^{*}}^{3}+3z_{d_{i}^{*}}^{3}+2z_{L_{i}}^{3}+z_{e_{i}^{*}}^{3}+z_{\nu_{i}^{*}}^{3}),\quad \left[U(1)^{\prime}{}^{3}\right].\label{eq:1p-1p-1p}
\end{eqnarray}
Here $i$ is the family index.

These equations are non-linear and have too many free parameters. To
solve these equations, some assumptions are usually made, for instance, family universal charges assignment. In family universal models, anomaly cancelation is separately satisfied in each family. However, family universality is not a necessary condition, and more generally, there can exist family non-universal models as we show in the follows.

From the first four linear equations, we get three equations for the quark sector
\begin{eqnarray}
\sum_{i=1}^{3}z_{d_{i}^{*}} & = & \sum_{i=1}^{3}\left(\frac{4}{3}z_{L_{i}}+z_{e_{i}^{*}}\right),\nonumber \\
\sum_{i=1}^{3}z_{u_{i}^{*}} & = & -\sum_{i=1}^{3}(\frac{2}{3}z_{L_{i}}+z_{e_{i}^{*}}),\nonumber \\
\sum_{i=1}^{3}z_{Q_{i}} & = & -\frac{1}{3}\sum_{i=1}^{3}z_{L_{i}},
\label{eq:Anomalies4}
\end{eqnarray}
and one for the neutrino sector
\[
\sum_{i=1}^{3}z_{\nu_{i}^{*}}  =
-\sum_{i=1}^{3}\left(2z_{L_{i}}+z_{e_{i}^{*}}\right).
\]
Right-handed neutrinos are introduced here since we now know
neutrinos have masses but whether they are Dirac or Majorana particles is still unknown. Here and after, we will focus on the quark sector and not discuss neutrinos any further.

\section{realistic $Z'$ boson model}

The LEPII and hadron colliders have put stringent limits on the $Z'$ boson of popular models \cite{Langacker:2008yv}, including its mass and couplings to leptons as well as the mixing angle with $Z$, by
using mostly the lepton decay channel of $Z'$ boson. To fit the experimental
results and escape the limits, an usual assumption made in the literature is that the
$Z'$ couplings with the leptons are much smaller than those with
quarks, namely the U(1)$'$ charges of leptons are much smaller than those of
quarks. Those models are called leptophobic $Z'$ models. Actually, leptophobic theory is not so strange as we do have a similar structure in standard model, i.e., SU(3)$_c$ QCD of strong interaction.

\subsection{Anomaly cancelling separately in quark and lepton sectors}

In our present considerations, we will not impose leptophobic structure as a
constraint in our model, but treat the U(1)$'$ charges of quarks and leptons
to be independent with the only constraint from the anomaly cancelation. From eq.~(\ref{eq:Anomalies4}), the U(1)$'$ charges of leptons and quarks should naturally be at the same order. If there is a hierarchy in charges of lepton and quark sectors, there should be, among their charges, some extra or hidden relations which are beyond the scope of our discussions in this paper.

It is interesting to notice that in the linear equations given in
eq.~(\ref{eq:Anomalies4}),  when both sides of the equations are zero,
the cancelation of U(1)$'$ anomalies may be satisfied separately in lepton and
quark sectors among three families, based on their similar structures in the last two non-linear equations of eqs.~(\ref{eq:1p-1p-1p}). In that case, the charges of quarks are unrelated to those of leptons, and then
the equations for quarks read
\begin{eqnarray}
0 & = & \sum_{i=1}^{3}z_{Q_{i}},0=\sum_{i=1}^{3}z_{u_{i}^{*}},0=\sum_{i=1}^{3}z_{d_{i}^{*}},\nonumber \\
0 & = & \sum_{i=1}^{3}(z_{Q_{i}}^{2}-2z_{u_{i}^{*}}^{2}+z_{d_{i}^{*}}^{2}),\nonumber \\
0 & = &
\sum_{i=1}^{3}(2z_{Q_{i}}^{3}+z_{u_{i}^{*}}^{3}+z_{d_{i}^{*}}^{3}).\label{eq:Assum1}
\end{eqnarray}
and similar equations for leptons,
\begin{eqnarray*}
0 & = & \sum_{i=1}^{3}z_{L_{i}},\;0=\sum_{i=1}^{3}z_{e_{i}^{*}},\;0=\sum_{i=1}^{3}z_{\nu_{i}^{*}},\\
0 & = & \sum_{i=1}^{3}(-z_{L_{i}}^{2}+z_{e_{i}^{*}}^{2}),\\
0 & = &
\sum_{i=1}^{3}(2z_{L_{i}}^{3}+z_{e_{i}^{*}}^{3}+z_{\nu_{i}^{*}}^{3}).\label{eq:Assum2}
\end{eqnarray*}
Note that the sum over family indices, and that if
anomaly cancelation occurs in each family, then all charges are zero, a trivial assignment.

\subsection{Family Universality for the First and Second Families}

Family non-universality generally leads to flavor mixing among fermions in their
interactions with $Z'$ boson in the mass eigenstates. In order to suppress the FCNC effects to fit the experimental results, especially the FCNC in the first two families are strongly constrained from the kaon meson system, thus we consider U$(1)'$ charges of the first and second families to be the same,
\[
z_{Q_{1}}=z_{Q_{2}},\; z_{u_{1}^{*}}=z_{u_{2}^{*}},\;
z_{d_{1}^{*}}=z_{d_{2}^{*}}.
\]
With this assumption, the last two equations given in Eq.~(\ref{eq:Assum1})
can be written as
\begin{eqnarray*}
\sum_{i=1}^{3}(z_{Q_{i}}^{2}-2z_{u_{i}^{*}}^{2}+z_{d_{i}^{*}}^{2}) & =0\Rightarrow & z_{Q_{1}}^{2}-2z_{u_{1}^{*}}^{2}+z_{d_{1}^{*}}^{2}=0,\\
\sum_{i=1}^{3}(2z_{Q_{i}}^{3}+z_{u_{i}^{*}}^{3}+z_{d_{i}^{*}}^{3}) &
=0\Rightarrow &
2z_{Q_{1}}^{3}+z_{u_{1}^{*}}^{3}+z_{d_{1}^{*}}^{3}=0.
\end{eqnarray*}
The above equations have a very simple solution
\[
z_{u_{1}^{*}}=z_{d_{1}^{*}}=-z_{Q_{1}}\equiv-z.
\]
Although this is not the only solution, while other five solutions are
complex, irrational and complicated. We shall restrict ourselves to
the above simple solution in our present consideration. We can rescale the coupling
$g_{1}^{'}$ so that $z=1$ for later convenience.

\begin{table}[tbph]
 \begin{tabular}{|c|c|c|c|c|}
\hline
 & SU(3)$_{c}$  & SU(2)$_{L}$  & U(1)$_{Y}$  & U(1)$^{\prime}$ \tabularnewline
\hline $Q_{L}^{i}$  & $3$  & $2$  & $+1/3$  &
$\left(+1,+1,-2\right)$\tabularnewline \hline $u_{R}^{i*}$  &
$\bar{3}$  & $1$  & $-4/3$  & $\left(-1,-1,+2\right)$\tabularnewline
\hline $d_{R}^{i*}$  & $\bar{3}$  & $1$  & $+2/3$  &
$\left(-1,-1,+2\right)$\tabularnewline \hline
\end{tabular}
\caption{U$(1)'$ charges of quarks.}
\end{table}

For completeness, we also list the U$(1)'$ charges of the leptons. It is
straightforward to check that the following assignment of the U(1)$'$
charges can satisfy the above equations, up to a constant $\mu$
relative to the quark sector. %
\begin{table}[tbph]
 \begin{tabular}{|c|c|c|c|c|}
\hline
 & SU(3)$_{c}$  & SU(2)$_{L}$  & U(1)$_{Y}$  & U(1)$^{\prime}$ \tabularnewline
\hline $L^{i}$  & $1$  & $2$  & $-1$  &
$\mu\left(+1,+1,-2\right)$\tabularnewline \hline $e_{R}^{i*}$  & $1$
& $1$  & $+2$  & $\mu\left(-1,-1,+2\right)$\tabularnewline \hline
$\nu_{R}^{i*}$  & $1$  & $1$  & $0$  &
$\mu\left(-1,-1,+2\right)$\tabularnewline \hline
\end{tabular}
\caption{U$(1)'$ charges of leptons.}
\end{table}
Here we will leave $\mu$ as an arbitrary constant. When $\mu \ll 1$,
it becomes a leptophobic model. And $\mu \approx 1$ indicates that the U(1)$'$ charges of
leptons and quarks are at the same order.

An important feature in this model is that the U(1)$'$
charges of the third family are as twice large as those of the first
and second families. This feature reduces the branching radio of $Z' \rightarrow
l^-l^+(l=e,\mu)$ in dilepton search, and then the constraints may also be relaxed correspondingly. In the later discussion of experiment constraint, we will consider the limits from both the dijet and dilepton search.

\subsection{Yukawa and Higgs sectors}

Based on the U$(1)'$ charges of fermions, the SM Higgs
doublet $H_{1}$ with zero $U(1)'$ charge can cause the spontaneous symmetry breaking to generate the masses of all the SM particles. The Yukawa interactions are
\begin{eqnarray}
\mathcal{L}_{H_{1}}&=&\sum_{i,j=1}^{2}f_{ij}^{u}\bar{Q}_{L,i}\tilde{H}_{1}u_{R,j}+f_{33}^{u}\bar{Q}_{L,3}\tilde{H}_{1}u_{R,3}\nonumber\\
&+&\sum_{i,j=1}^{2}f_{ij}^{d}\bar{Q}_{L,i}H_{1}d_{R,j}+f_{33}^{d}\bar{Q}_{L,3}H_{1}d_{R,3}+h.c\label{eq:Y_H1}
\end{eqnarray}
The resulting two mass matrices have the following forms
\[
\mathcal{M}_{u,d}^{H_{1}}\sim\left(\begin{array}{ccc}
\times & \times & 0\\
\times & \times & 0\\
0 & 0 & \times\end{array}\right).
\]
which shows that when the Yukawa term eq.(\ref{eq:Y_H1}) is the only interaction, it cannot obtain the realistic cabbibo-Kobayashi-Maskawa(CKM) matrix for quark mixing and CP violation. Thus we need to introduce two more Higgs doublets with non-zero U$(1)'$ charges. Specifically taking $H_{2}$ with U(1)$'$ charge $-3$ and $H_{3}$ with $+3$, we have the additional Yukawa interactions for $H_{2}$ and $H_{3}$,
\begin{eqnarray}
\mathcal{L}_{H_{23}}
& = & f_{13}^{u}\bar{Q}_{L,1}\tilde{H}_{2}u_{R,3}+f_{23}^{u}\bar{Q}_{L,2}\tilde{H}_{2}u_{R,3}\nonumber \\
& + & f_{31}^{u}\bar{Q}_{L,3}\tilde{H}_{3}u_{R,1}+f_{32}^{u}\bar{Q}_{L,3}\tilde{H}_{3}u_{R,2}\nonumber \\
& + & f_{13}^{d}\bar{Q}_{L,1}H_{3}d_{R,3}+f_{23}^{d}\bar{Q}_{L,2}H_{3}d_{R,3}\nonumber \\
& + &
f_{31}^{d}\bar{Q}_{L,3}H_{2}d_{R,1}+f_{32}^{d}\bar{Q}_{L,3}H_{2}d_{R,2}+h.c.
\label{eq:Y_H23}
\end{eqnarray}
These Yukawa interactions will contribute to the mass matrices with\[
\mathcal{M}_{u,d}^{H_{23}}\sim\left(\begin{array}{ccc}
0 & 0 & \times\\
0 & 0 & \times\\
\times & \times & 0\end{array}\right).\]
Similarly, one can write down the Yukawa interactions in the
lepton section. Note that for $\mu\neq\pm1$, two extra Higgs doublets have to be introduced.

When $H_2$ and $H_3$ get vacuum expectation values(VEVs), the U$(1)'$ gauge symmetry will also be broken as they carry nonzero U(1)$'$ charges. If the U$(1)'$ gauge coupling is at the same order as the electroweak coupling, the mass
of $Z'$ boson is then at the same order as the electroweak scale, which is highly constrained. To get a heavy $Z'$ boson, we need to introduce a singlet scalar  $S$ with U(1)$'$ charge. With three Higgs doublets and their U(1)$'$ charges as well as a singlet scalar (its U(1) charge $z_{s}$ is unspecified), the most general gauge invariant Higgs potential is given by
\begin{eqnarray}
V_{H}
& = & -m_{1}^{2}|H_{1}|^{2}-m_{2}^{2}|H_{2}|^{2}-m_{3}^{2}|H_{3}|^{2}-m_{S}^{2}|S|^{2}\nonumber \\
& & +\lambda_{1}|H_{1}|^{4}+\lambda_{2}|H_{2}|^{4}+\lambda_{3}|H_{3}|^{4}+\lambda_{S}|S|^{4}\nonumber \\
& & +\lambda_{4}|H_{1}|^{2}|H_{2}|^{2} + \lambda_{5}|H_{1}|^{2}|H_{3}|^{2} + \lambda_{6}|H_{2}|^{2}|H_{3}|^{2} \nonumber \\
& & +\lambda_{7}(H_{1}^{\dagger}H_{2})(H_{2}^{\dagger}H_{1}) + \lambda_{8}(H_{1}^{\dagger}H_{3})(H_{3}^{\dagger}H_{1})+ \lambda_{9}(H_{2}^{\dagger}H_{3})(H_{3}^{\dagger}H_{2})\nonumber \\
& &+\left[\lambda_{1s}|H_{1}|^{2}+\lambda_{2s}|H_{2}|^{2}+\lambda_{3s}|H_{3}|^{2}\right]|S|^{2}\nonumber \\
&
&+\left(\lambda_{10}H_{1}^{\dagger}H_{2}H_{1}^{\dagger}H_{3}+h.c\right)\label{eq:HiggsV}
\end{eqnarray}
where all parameters $m_{i}$ and $\lambda_{i}$ are real except that
the parameter $\lambda_{10}=|\lambda_{10}|e^{i\alpha}$ could be complex
potentially. There may exist terms like
$H_{2}^{\dagger}H_{3}S^{n}(n\leq2)$ when choosing the U(1)$'$ charge $z_{s}=-6$ for $n=1$ and $z_s=-3$ for $n=2$. Without specifying the charge $z_{s}$, we may not discuss such terms in the rest of
our paper. Considering the following vacuum structure which breaks the gauge symmetries
to the electromagnetic symmetry U$(1)_Q$,
\begin{equation}
\langle H_{i}\rangle=\left(\begin{array}{c}
0\\
\frac{v_{i}}{\sqrt{2}}e^{i\theta_{i}}\end{array}\right),\; i= 1,2,3; \qquad \langle S\rangle=v_{s}/\sqrt{2}
\label{eq:VEV}
\end{equation}
One can always use the global U$(1)$ symmetry to make $\theta_{1}=0$. In eq.(\ref{eq:HiggsV}), the only term that depends on
the phase is $\lambda_{10}H_{1}^{\dagger}H_{2}H_{1}^{\dagger}H_{3}$, when minimalizing the Higgs potential for the complex field component of the Higgs doublet $H_1$, we are led to the following minimal condition
\begin{equation}
\sin\left(\alpha+\theta_{2}+\theta_{3}\right) = 0,\qquad \alpha = -(\theta_{2}+\theta_{3})
\end{equation}
which indicates that $\alpha$ is tightly related to the vacuum phases, so CP symmetry is broken via the vacuum though it is not exactly spontaneous CP violation.

\textit{Spontaneously CP violation}: When $\lambda_{10}$ is real,
namely, $\alpha=0\text{ or }\pi$ and $\theta_2=-\theta_3$, there may exist some vacuum
configurations that can cause spontaneous $CP$ violation in such
a three Higgs doublet model. As there is an Abelian gauge symmetry,
the form of the Higgs potential is constrained. For a
general discussion on those models, it is referred to \cite{Branco:1983tn}. When Yukawa
interactions are not considered, it is possible to redefine the
$CP$ transformations of the Higgs fields, so that no spontaneous $CP$
violation occurs in the Higgs sector. With Yukawa interactions of fermions, such
redefinitions will induce complex Yukawa couplings.

\textit{Masses of gauge bosons}: It is known that the gauge bosons will get their masses from their
interactions with Higgs bosons after spontaneous symmetry breaking.
\[
\mathcal{L}_{H}=\sum_{i=1}^{3}\left(D^{\mu}H_{i}\right)^{\dagger}\left(D_{\mu}H_{i}\right)+\left(D^{\mu}S\right)^{\dagger}\left(D_{\mu}S\right)
\]
the covariant differential operators can be easily written down and
we omit them here. The mass term for the charged gauge bosons is simply given by
\begin{eqnarray*}
\frac{1}{4}g_{2}^{2}\left[v_{1}^{2}+v_{2}^{2}+v_{3}^{2}\right]W_{\mu}^{-}W^{+\mu}.
\end{eqnarray*}
For the neutral gauge bosons, their mass terms have the following forms
\begin{eqnarray*}
 &  & \frac{v_{1}^{2}}{8}g^{2}Z_{0}^{\mu}Z_{0\mu}+\frac{v_{S}^{2}}{8}g_{1}^{'2}z_{s}^{'2}B_{\mu}^{'}B^{'\mu}\\
 & + & \frac{v_{2}^{2}}{8}\left(gZ_{0\mu}-g_{1}^{'}z_{2}^{'}B_{\mu}^{'}\right)\left(gZ_{0}^{\mu}-g_{1}^{'}z_{2}^{'}B^{'\mu}\right)\\
 & + & \frac{v_{3}^{2}}{8}\left(gZ_{0\mu}-g_{1}^{'}z_{3}^{'}B_{\mu}^{'}\right)\left(gZ_{0\mu}-g_{1}^{'}z_{3}^{'}B^{'\mu}\right)
\end{eqnarray*}
with $z'_2$ and $z'_3$ the U(1)$'$ charges of the Higgs doublets $H_2$ and $H_3$ respectively, and $z'_2 = -z'_3=-z'_h$ in our model. Where we have used the definition for the SM gauge bosons
\begin{eqnarray*}
Z_{0\mu} & = & \frac{1}{\sqrt{g_{1}^{2}+g_{2}^{2}}}\left[g_{2}W_{\mu}^{3}-g_{1}B_{\mu}\right]=\cos\theta_{W}W_{\mu}^{3}-\sin\theta_{W}B_{\mu},\\
A_{\mu} & = &
\frac{1}{\sqrt{g_{1}^{2}+g_{2}^{2}}}\left[g_{1}W_{\mu}^{3}+g_{2}B_{\mu}\right]=\sin\theta_{W}W_{\mu}^{3}+\cos\theta_{W}B_{\mu}.
\end{eqnarray*}
Here $A_{\mu}$ is the photon field and $g^{2}=g_{1}^{2}+g_{2}^{2}$.
The mass matrix for ($Z_{0\mu},B_{\mu}^{'}$) is found to be
\begin{eqnarray}
M^2 &= &
\frac{1}{8}\left(\begin{array}{cc}
g^{2}\left[\sum_{i=1}^{3}v_{i}^{2}\right] & gg_{1}^{'}z'_h \left[v_{2}^{2}-v_{3}^{2}\right]\\
gg_{1}^{'}z'_h\left[v_{2}^{2}- v_{3}^{2}\right] &
g_{1}^{'2}\left[(v_{2}^{2}+v_3^2) z_{h}^{'2}+v_{s}^{2}z_{s}^{'2}\right]\end{array}\right)  \nonumber
\end{eqnarray}
Note that the off-diagonal terms can be made small when $v_{2}\approx v_{3}$, so that there is a negligible mixing between $Z_{\mu}$$\left(Z_{\mu}=Z_{0\mu}\right)$ and $Z_{\mu}^{'}\left(Z_{\mu}^{'}=B_{\mu}^{'}\right)$.  Generally, it happens
when imposing a permutation symmetry on the two Higgs doublets $H_2$ and $H_3$ under $H_{2}\leftrightarrow
H_{3}$, and in that case the off-diagonal terms vanish. While such a symmetry can not be an
exact, otherwise the form of Yukawa terms will be highly constrained. In the spirit of simplicity for later phenomenological constraints, we will impose $v_{2}\approx v_{3}$ and then experimental constraints from $Z-Z'$ mixing can be relaxed.

\section{Phenomenological Constraints of $Z'$ Boson}

We are now ready to consider in our model the phenomenological effects of $Z'$ boson and focus mostly on the quark sector to constrain the parameters, $M_{Z'}$ and $g_{1}^{'}$. We shall show that the $Z'$ boson in this model, generally, allow both flavor conserving and flavor changing couplings at the mass eigenstates of quarks. In the flavor conserving processes, we use the latest dijet search result to constrain the production rate of $Z'$ and give some limits on $M_{Z'}$ and $g_{1}^{'}$.  While in the flavor changing processes, we are going to investigate the t-channel $Z'$ effects with only paying attention to the right-handed up quark part.

\subsection{Discussion framework}

The Lagrangian for $\bar{q}qZ'$ couplings is
\begin{equation}
\mathcal{L}_{F}=\sum_{i=1}^{3}\left[\bar{Q}_{L}^{i}i\gamma^{\mu}D_{\mu}Q_{L}^{i}+\bar{u}_{R}^{i}i\gamma^{\mu}D_{\mu}u_{R}^{i}
+\bar{d}_{R}^{i}i\gamma^{\mu}D_{\mu}d_{R}^{i}\right]\label{eq:f_kinetic}
\end{equation}
where the covariant differential operators on the left-handed and
right-handed fermions are implicitly implied. We may rewrite interactions of gauge boson $Z^{'}$ in the
gauge eigenstates as $g_{1}^{'}Z_{\mu}^{'}J_{Z^{'}}^{\mu}$,
\[
J_{Z^{'}}^{\mu}=\sum_{\psi}\sum_{i=1}^{3}\bar{\psi}_{i}\gamma^{\mu}\left[\epsilon_{i}^{\psi_{L}}P_{L}+\epsilon_{i}^{\psi_{R}}P_{R}\right]\psi_{i}\;,\;\psi=u,d
\]
where the conventions are borrowed from \cite{Barger:2009qs}. After the
gauge symmetries are broken down spontaneously, we need to diagonalize the mass
matrices by redefining the quark fields,
\begin{eqnarray}
u_{R}^{i} & = & \left(V_{U_{R}}\right)_{ij}U_{R}^{j},\;
u_{L}^{i}=\left(V_{U_{L}}\right)_{ij}U_{L}^{j},\nonumber \\
d_{R}^{i} & = & \left(V_{D_{R}}\right)_{ij}D_{R}^{j},\;
d_{L}^{i}=\left(V_{D_{L}}\right)_{ij}D_{L}^{j}.
\label{eq:f_rotation}
\end{eqnarray}
The CKM matrix is given by
 \begin{equation}
V_{\text{CKM}}=V_{U_{L}}^{\dagger}V_{D_{L}}.\label{eq:CKM}
\end{equation}
The transformation on the fermion fields Eq.(\ref{eq:f_rotation}) leads to
\[
J_{Z^{'}}^{\mu}=\sum_{\psi=(U,D)}\sum_{i,j=1}^{3}\bar{\psi}_{i}\gamma^{\mu}\left[V_{\psi_{L}}^{\dagger}\epsilon^{\psi_{L}}V_{\psi_{L}}P_{L}+V_{\psi_{R}}^{\dagger}\epsilon^{\psi_{R}}V_{\psi_{R}}P_{R}\right]\psi_{j}.
\]
The above interactions can give rise to FCNC effects when $V^{\dagger}\epsilon
V\not\propto I$, which occurs when the $Z^{'}$ charges are not family
universal, i.e., $\epsilon\not\propto I$. In our present consideration, the matrix $\epsilon$ is
universal for the left and right handed fermions, and also for the up and down quarks in the first and second
families,
\[
\epsilon^{\psi_{L,R}}=\left(\begin{array}{ccc}
+1\\
 & +1\\
 &  & -2\end{array}\right).
 \]
Since only the combination $V_{\text{CKM}}=V_{U_{L}}^{\dagger}V_{D_{L}}$ is experimentally
known, and the individual matrix $V_{\psi_{L,R}}$ are unknown,  then
the resulting $B^{\psi_{L,R}}\equiv V_{\psi_{L,R}}^{\dagger}\epsilon^{\psi_{L,R}}V_{\psi_{L,R}}$ is also
unknown.  One may make a rotation such that the coupling matrices have the following approximate form
\begin{equation}
B^{\psi_{L,R}}\simeq\left(\begin{array}{ccc}
B_{11}^{\psi_{L,R}} & 0 & B_{13}^{\psi_{L,R}}\\
0 & B_{22}^{\psi_{L,R}} & B_{23}^{\psi_{L,R}}\\
B_{13}^{\psi_{L,R}*} & B_{23}^{\psi_{L,R}*} &
B_{33}^{\psi_{L,R}}\end{array}\right). \label{eq:B-matrice}
\end{equation}
which fits to the phenomenology of light and heavy meson systems. More general discussions on the B meson physics with the above form of couplings have been given in \cite{Barger:2009qs}. For the down-quark sector, \cite{Barger:2009qs} demonstrates that such family non-universal $U(1)'$ scenarios can account for the
currently observed discrepancies with the SM predictions for $B_{s}-\bar{B}_{s}$ mixing and the
time-dependent CP asymmetries of the penguin-dominated $B_{d}\rightarrow (\pi, \phi, \eta', \rho,\omega, f_{0})K_{S}$
decays. In the present paper, we shall discuss mainly on the up-quark sector and consider some effects on top quark
observables.

We would like to point out that although
$\epsilon^{\psi_{L}}=\epsilon^{\psi_{R}}$, it is not necessary to indicate
that $B_{ij}^{\psi_{L}}=B_{ij}^{\psi_{R}}$ as $V_{\psi_{L}}$ and
$V_{\psi_{R}}$ could generally be different. When $B_{ij}^{\psi_{L}} \neq B_{ij}^{\psi_{R}}$, there will be observable effect, such as forward-backward asymmetry. As the coupling matrices can have
both diagonal and off-diagonal non-zero elements, there are also both flavor conserving
and flavor changing phenomenology via $Z'$ boson interactions.

\subsection{Constraint from dijet search}

\begin{figure}[t]
\includegraphics[scale=0.35]{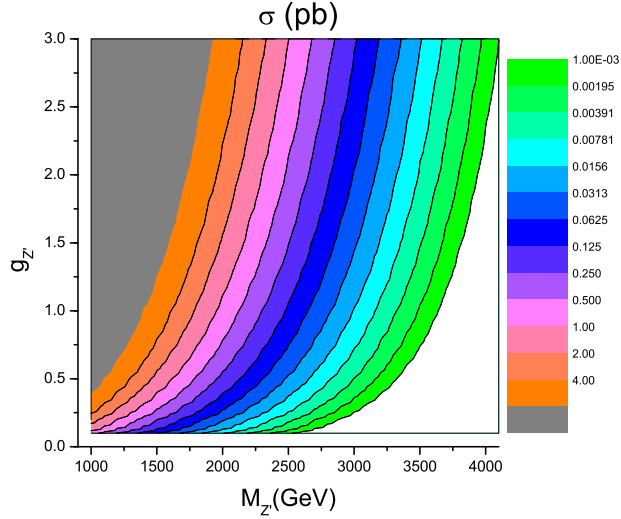}
\caption{Production rate or cross section ($pp \rightarrow Z'$) as contours of $M_{Z'}$ and gauge coupling $g_{Z'}$ at the LHC with $\sqrt{s}=7$ TeV.} \label{fig:MGcontour}
\end{figure}
\begin{figure}[t]
\includegraphics[scale=0.35]{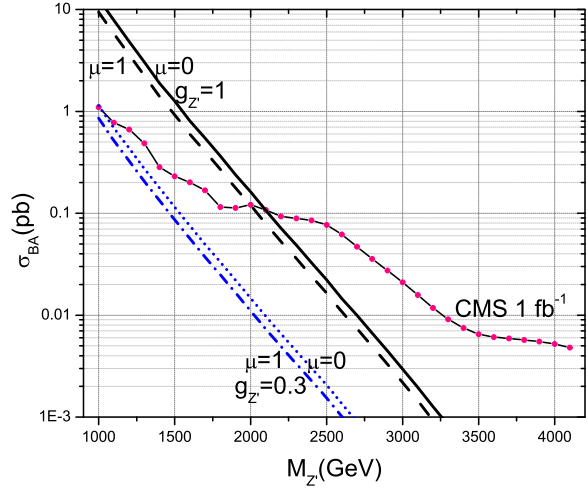}
\caption{Contraints from dijet search at the CMS(1.0 fb$^{-1}$) \cite{CMSdijet}. Black solid(dashed) line denotes the $\mu=0(\mu = 1)$ with $g_{Z'}=1$, and blue dot(dot-dashed) line represents the $\mu=0(\mu = 1)$ with $g_{Z'}=0.2$. Red-point line indicates the $95 \%$ CL upper limit on $\sigma \times B\times A$. See text for details. }
\label{fig:CMSZp}
\end{figure}

In the subsection, we will focus on the flavor-conserving components and meanwhile treat the flavor-changing ones of $q_i\bar{q}_jZ^{'}$ coupling matrix vanishing. We can then write the coupling matrix as, $G_{ij}\equiv V_{\psi_{L}}^{\dagger}\epsilon^{\psi_{L}}V_{\psi_{L}}P_{L}+V_{\psi_{R}}^{\dagger}\epsilon^{\psi_{R}}V_{\psi_{R}}P_{R}= \epsilon \delta_{ij}$. Coupling of this form can be achieved by properly adjusting $V_{\psi_{L}}$ and $V_{\psi_{R}}$.

We mainly focus on the TeV scale $Z'$ in this sector. In Fig.\ref{fig:MGcontour}, we plot a color map of the cross section($pp\rightarrow Z'$) at the LHC with $\sqrt{s}=7$ TeV as functions of $M_{Z'}$ and gauge coupling $g_{Z'}$. At the LHC, $Z'$ is produced through the channel $q\bar{q}\rightarrow Z'$ and then decay to quark pairs and$/$or lepton pairs, depending on $\mu$, the ratio of quarks and lepton's charge. Search for s-channel resonance can put lower bound on the mass $M_{Z'}$ and correspondingly upper bound on gauge couplings $g_{Z'}$.

The dijet searches \cite{CMSdijet,ATLASdijet} are directly connected with s-resonance $Z'$.  It can constrain this U$(1)'$ model as shown in Fig.\ref{fig:CMSZp}. We use the result from CMS Collaboration \cite{CMSdijet}. The result can be easily compared with the total cross section($\sigma$) at parton level and effective cross section($\sigma_{BA}$). Here $\sigma_{BA}$ stands for $\sigma \times B\times A$, where $B$ for branching fraction to dijet and $A$ for acceptance. The value of $A\sim 0.6$, the same one used in CMS Collaboration \cite{CMSdijet}, is independent of the mass of $Z'$. The branching ratio $B$ is different for $\mu=1$ and $\mu=0$(leptophobic model). For $M_{Z'}$ of TeV scale, quarks and leptons' masses can be neglected and straightforward calculation gives $B=1/4$ for $\mu=1$ and $B=1/3$ for $\mu=0$. Correspondingly, the upper limits are different in two cases(for $\mu=1$, constraint from dilepton search need to be considered, as discussed below). As we can see from the Fig.(\ref{fig:CMSZp}), the exclusion limit for $M_{Z'}$ is slight changed, up to $\mathcal{O}(100)$ GeV. As the constraints  depends on the gauge couplings $g_{Z'}$, we plot for two values of $g_{Z'}$. For $g_{Z'}=1$, the figure shows that $M_{Z'}<2.1$ TeV is excluded at $95\%$ confidence level(CL), while for $g_{Z'}=0.3$ the limit is lowered to $1.0$ TeV.

\subsection{Constraint from dilepton search}

For $\mu = 0$, there is no constraint from leptonic processes since no interaction exists between $Z'$ and leptons. In principle, for $\mu \neq 0$, in addition to the constraint from dijet search, we also need combine with the limit from dilepton search \cite{CMSdilepton,ATLASdilepton}, $Z'\rightarrow l^{-}l^{+}(l=e,\mu)$. Since there are only exclusion figures but no accessible and quantitative numbers, we can not repeat the same process as dijet case, but only give some qualitative analysis by comparing with sequential standard model(SSM).

For standard-model-like $Z^{'}_{\textrm{SSM}}$, with $g^u_{V}=0.29, g^u_{A}=0.50, g^d_{V}=-0.33, g^d_{A}=0.52$ and Br$(Z^{'}_{\textrm{SSM}}\rightarrow l^-l^+)=6.73\%$, CMS experiment \cite{CMSdilepton} has excluded at $95\%$ confidence level the mass of $Z^{'}_{\textrm{SSM}}$, $M_{Z'}>1940$ GeV. For our model, several differences can modify the limit. The branching ratio of $Z'\rightarrow l^+l^-$, Br$(Z'\rightarrow l^-l^+)$ is quadratically dependent on the $\mu$ with $B_{l^-l^+} \propto \mu^2 \times \frac{1}{4} \times \frac{1}{6}$. The factor $\mu^2/4$ arises because other $3/4$ of $Z'$ decay to quarks for $\mu=1$ and the factor $\frac{1}{6}$ originates the fact that,  of the leptonic decaying $Z'$, half of them decay to neutrinos and $\frac{1}{3}$ to $\tau$s. So the ratio of production rate reads
\begin{equation}
R_\sigma=\frac{\sigma(pp\rightarrow Z'\rightarrow l^+l^- )}{\sigma(pp\rightarrow Z^{'}_{\textrm{SSM}} \rightarrow l^+l^- )} \simeq \frac{4g^2_{Z'}}{{g^{u}_{V}}^2+{g^{u}_{A}}^2+{g^{d}_{V}}^2+{g^{d}_{A}}^2}\times\frac{\mu^2/24}{0.0673}.
\end{equation}
We can use the above quantity to estimate the bounds for $M_{Z'}$ for different $g_{Z'}$. For $\mu = 1 $ and $ g_{Z'}=0.3$, $R_\sigma \simeq\frac{5}{16}$, and this will change the exclusion limit of $M_{Z'}$ to $1.6$ TeV by comparing the figure presented in \cite{CMSdilepton}. For $\mu = 1/2$ and $ g_{Z'}=0.3$, the limit can be shifted to $1.25$ TeV. In both case, limits from dilepton search are more stringent than those from dijet search as expected, since leptonic final states have better discriminating power.

\subsection{The t-channel $Z'$ effects}
As mentioned above, this model also allows flavor-changing and left-right asymmetry couplings. In such a case, the limits from s-channel dijet or dilepton searches can not be literally applied. The trends show that the constraints on the $M_{Z'}$ and $g_{Z'}$ are correlated, the smaller $g_{Z'}$, the more loosely constrained $M_{Z'}$. We also know that B physics can give stringent constraint on the coupling of down quark sector. So in the section, we shall only consider some effects on top quark sector within the lower $M_{Z'}$ region.

Both CDF and D0 have observed forward-backward asymmetry of top pair production at the Tevatron \cite{Abazov:2007qb,
Aaltonen:2008hc, CDFnote10436, CDFnote10185, Aaltonen:2011kc, Abazov:2011rq}. The CDF experiment has reported the forward-backward asymmetry at the parton-level \cite{CDFnote10185,Aaltonen:2011kc}(We mostly compare with the CDF result since it is at parton level which can be directly compared with theoretical calculation.),
\[
A^{t\bar{t}} = 0.158\pm 0.072\pm 0.017.
\]
In the SM with NLO-QCD corrections, result is given by
\[
A_{SM}^{t\bar{t}} = 0.058\pm 0.009.
\]
Even within the electroweak correction \cite{Hollik:2011ps}, standard model still can not give such a large asymmetry. The CDF experiment has also measured the total cross section $\sigma_{t\bar{t}}=7.50\pm 0.48\;$pb \cite{CDFnote9913} for top quark mass $m_{t}=172.5$ GeV, which is in agreement with the theoretical prediction in the SM
$\sigma_{t\bar{t}}=7.46_{-0.80}^{+0.66}\;$pb \cite{Langenfeld:2009wd, AguilarSaavedra:2011vw, Gresham:2011pa, Dorsner:2009mq}. The agreement of the cross section motivates light flavor changing $Z'$ effects. There are many related works on discussing this effect \cite{Haisch:2011up, AguilarSaavedra:2011ug, Gresham:2011fx, Jung:2011id, KSZ, Bai:2011ed, Jung:2009jz, M. R. Buckley, Berger:2011ua}. In the following, we shall give a short discuss in our present model.

In our model, when the coupling between u and t quark, $g^{L,R}_{ut}$ are not vanishing and $g^R_{ut}\neq g^L_{ut}$, it can lead to forward-backward asymmetry of top pair production and same top production at hadron collider. In this subsection, we analyse in our model such effects and in addition constrain the parameters. Our conclusion is that such a model can not account for all the forward-backward asymmetry observed at Tevatron if satisfying the limit from the same top production at the LHC \cite{Chatrchyan:2011dk}.

\begin{figure}[htb]
\includegraphics[scale=0.35]{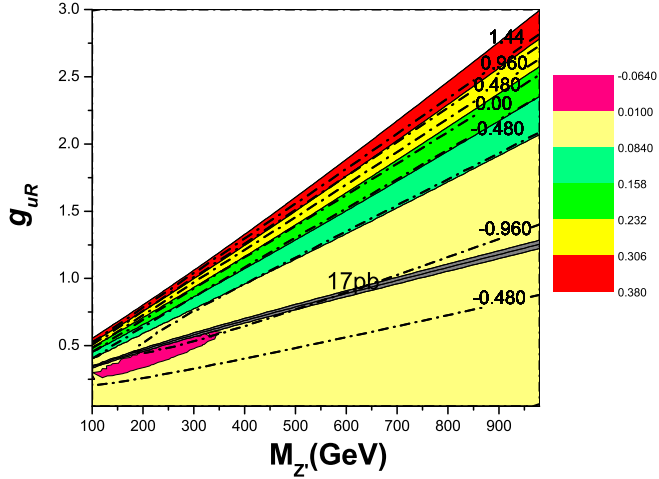}
\caption{Colored map for forward-backward asymmetry as function of
$g_{\rm{uR}}$ and $M_{Z'}(\textrm{GeV})$. The dot-dashed lines are
contours for the total cross section from NP. The dark-grey band
indicates the limit on the cross section of same top production at
the LHC. We allow $10\%$ variance to make the band easy to see.
Region above the band gives cross section larger than 17pb. }
\label{fig:AFB1000}
\end{figure}

For simplicity, we may only consider the typical
case $B_{13}^{U_L}=0$ and give an upper bound on $g_{uR}\equiv g^R_{ut}$. In Fig.~\ref{fig:AFB1000}, we show the
colored map of $A_{FB}$ in the possible range of parameters: 100 GeV $ < M_{Z'}< $ 1000 GeV and
$0.05<g_{uR}<3$. Such t-channel process will
also contribute to the total cross section of $t\bar{t}$ pair
production denoted by $\sigma_{\textrm{NP}}$(here NP labels New Physics), we plot
the contours of $\sigma_{\textrm{NP}}$ in dot-dashed lines in the
same figure.

When the t-channel $Z'$ boson exchange is responsible for the asymmetry, it will also
induce two top quarks or two anti-top quarks production. The experiments at Tevatron
have already constrained the cross section, while the experiment at LHC
\cite{Chatrchyan:2011dk} can give even more stringent constraint on the
coupling for the same sign top quark production $\left(uu\rightarrow
tt\right)$ through the t-channel or u-channel. In Fig.~\ref{fig:AFB1000}, we
indicate, in the $M_{Z'}-g_{uR}$ plot, with a grey band of 17 pb cross section for the $uu\rightarrow tt$ production reported by the LHC experiment. We allow 10\% uncertainty in order to make the band easily seen. The region above the grey band is disfavored, while the region below the band is allowed and the resulting $A_{FB}$ presents a reasonable prediction in our model.

\section{Conclusions}

We have presented a simple UV complete and family non-universal U$(1)'$ gauge model in which
the anomaly is canceled by appropriately choosing the U(1)$'$ charges of the SM particles without introducing extra fermions. As the anomaly is canceled separately in the lepton and quark sectors, which may leave a room to escape the strong constraints from the direct searches for the final states with two leptons. Also the U$(1)'$ charges of the first and second families are taken to be the same in order to suppress the FCNC effects. To build a realistic model via spontaneous symmetry breaking, it requires one Higgs doublet with zero U(1)$'$ charge and two Higgs doublets and a singlet scalar with non-zero U$(1)'$ charges. As the U$(1)'$ charges of the first and second families are different from those of third family, there exist in general both flavor conserving and flavor changing couplings of the $Z'$ boson after spontaneous symmetry breaking and rotating to the mass eigenstates of quarks and leptons.

In flavor-conserving case, we have used the latest dijet and dilepton search  to constrain the mass of $Z'$ and gauge coupling $g_{Z'}$. For a typical $g_{Z'}=0.3$, at $95\%$ confidence level, $M_{Z'}$ is large than $1.0$ TeV, and $2.1$ TeV for $g_{Z'}=1$. The constraints depend on whether such $Z'$ couples to leptons. The limits can be shifted to higher value, depending on the ratio of U(1)$'$ charges between quark and lepton sector, $\mu$. When $\mu=0$, this model has a leptophobic structure and lepton sector impose no constraint.

For flavor-change couplings, the $Z'$ boson can in general couple with left-handed and right-handed quarks and leptons differently, so it can contribute to the forward-backward asymmetry(FBA) of the top quark pair production at the Tevatron.  Such t-channel process with flavor changing $Z'$ boson exchange, in additon to the constraint from the CMS experiment on the cross section of the same top production, indicate that such a $Z'$ can not solely account for the FBA observed.

\vspace*{0.5cm}

\vspace{1 cm}

\centerline{{\bf Acknowledgement}}

\vspace{20 pt}

The authors would like to thank Y. F. Zhou for useful discussions. This work was supported in part by the
National Science Foundation of China (NSFC) under Grant \#No.
10821504, 10975170 and the key project of the Chinese Academy of Science.

\end{document}